\documentclass[namedreferences]{SolarPhysics}
\usepackage{epsfig}

\newcommand{\be}{\begin{equation}}
\newcommand{\ee}{\end{equation}}
\newcommand{\beq}{\begin{eqnarray}}
\newcommand{\eeq}{\end{eqnarray}}

\def\dj{d\kern-0.4em\char"16\kern-0.1em}

\begin{document}
\begin{article}
\begin{opening}

\title{The Position of High Frequency Waves with Respect to the Granulation Pattern}

\runningtitle{ Energy of high frequency waves in the low solar chromosphere}    

\author{ A.\surname{An\dj i\'c}$^{1}$}
\runningauthor{ A.~An\dj i\'c} 

\institute{$^{1}$ Astronomical Research Centre, Queen's University Belfast, University Road, Belfast, BT7 1NN, UK 
                  \email{a.andic@qub.ac.uk}\\
             }

\date{Received: date}


\begin{abstract}

High frequency velocity oscillations were observed in the spectral lines Fe {\sc i} $543.45$nm and $543.29$nm, using 2D spectroscopy with a Fabry- Perot and speckle reconstruction, at the VTT in Tenerife. We investigate the radial component of waves with frequencies in the range $8$ - $22$mHz in the internetwork, network and a pore. We find that the occurrence of waves do not show any preference on location and are equally distributed over down-flows and up-flows, regardless of the  activity of the observed area in the line of Fe {\sc i} $543.45$nm. The waves observed in the lower formed line of Fe {\sc i} $543.29$nm seem to appear preferentially over down-flows.

\end{abstract}
\end{opening}

\section{Introduction}
\label{intro}

According to Kalkofen (2001) the problem of chromospheric heating is a question of energy supply for the radiative emission. It is thought that high frequency acoustic oscillations may provide this energy supply. The generation of acoustic oscillations can be described by the 'Lighthill mechanism' (Lighthill, 1951; Proudman, 1952). Deubner (1983) argues that high frequency bursts are either generated by rising granules, or propagate more or less uniformly from deeper layers into the convection zone. \par
The frequencies of acoustic oscillations depend on various fluid flow parameters \cite{lighthill51}. The most energetic oscillations should be generated in those regions where the convective velocities are largest \cite{stein67}.\par
Kalkofen (1990) suggests that the location of acoustic oscillations depend on their frequency. In areas of strong magnetic fields at the cell boundary, Kalkofen calculates that heating is done by oscillations with frequencies of $1 - 3$mHz. The bright points are heated by oscillations with periods around $6$mHz; while areas free of magnetic fields will be heated by oscillations of higher frequencies.\par
Observations \cite{espagnet96} show that acoustic events occur preferentially in dark intergranular lanes, i.e. corresponding to down flows of plasma and conclude that the excitation of solar oscillations is associated with rapid cooling occurring in the upper convection layer. \par
 Dom\`\i nguez (2004), Andjic and Wiehr (2006), Andic and Vo\'cki\'c (2007) and Trujillo Bueno {\it et al.}(2004) have shown evidence for magnetic activity in the quiet sun. Domingues (2004) found that the magnetic flux in the quiet internetwork is smeared by insufficient resolution and therefore underestimated. He also found that there are areas in the quiet internetwork which have magnetic fields comparable to those from the pores. Trujillo Bueno {\it et al.}(2004) claim that we can only see $1$\% of the magnetic field present in quiet internetwork regions. Andjic and Wiehr (2006) and Andic and Vo\'cki\'c (2007) show evidence for some magnetic activity in the quiet internetwork regions used in this work.

\section{Observations}
\label{observations}

The instrument used in this work is the German Vacuum Tower telescope (VTT) on the Canary islands, with the Fabry-Perot spectrometer \cite{bendlin92}. The instrumental setup has been described in detail by Koschinsky {\it et al.} (2001). We have not used a polarimeter in this study. The observations were performed using the spectral lines of Fe {\sc i} at $543.45$nm and $543.29$nm in June 2004. The data sets used in this work are listed in Table \ref{tpodaci}. 

\begin{table}[ht]
\caption{ The data sets used in this work, dates, used lines and object of observation. Position $[0,0]$ corresponds to the Sun center.}
\label{tpodaci}       
\begin{tabular}{llllllll}
\hline\noalign{\smallskip}
Mark & Lines [nm]& Area& Coordinates& Images& Expo-& Cad-&Dur-\\
&&&&&sure&ence&ation\\
&&&&&[ms]&[s]&[min]\\ 
\noalign{\smallskip}\hline\noalign{\smallskip}
DS1 &543.45 \& 543.29& Quiet Sun&$[0,0]$ & 108 & 30 & 28.4&52.7\\
DS2 &543.45 \& 543.29 & Bright points&$[0,0]$ & 108 & 30 & 28.4&33.61\\
DS3 & 543.45 & Pore& $[0,0]$ &  108 & 30 & 28.3&23.19\\ 
DS4 & 543.45 & Quiet Sun& $[96.7,90.7]$ & 91 & 20 & 22.7&52.9\\ 
\noalign{\smallskip}\hline
\end{tabular}
\end{table}

\section{Data Reduction and Analysis Methods}
\label{data}

The procedures used for data reduction in this work are the same as those used in Andic (2007) and Andjic (2006). \par

The data reduction for the broad-band images includes the following procedures: subtraction of darks, flat fielding and speckle reconstruction. The reconstruction is performed with software developed by P. Sutterlin following the method of de Boer (1992, 1994). The program used for the reconstruction of narrow-band images was developed by Janssen (2003) and is based on the method developed by Keller {\it et al.} (1992). The velocity maps were calculated  for the line cores from the line bisectors using the technique described in detail by Andic (2007).  Line profiles were calculated using 3D radiative-hydrodynamical model of Asplund {\it et al.} (2000). This method is described in detail in the work of Shchukina and Trujillo Bueno (2001). The formation height in LTE is $308.2$km for Fe {\sc i} $543.29$nm and $660.2$km for Fe {\sc i} $543.45$nm. In NLTE the corresponding heights are $258.5$km and $588.7$km for Fe {\sc i} $543.29$nm and Fe {\sc i} $543.45$nm respectively (N. Shchukina, private correspondence).\par
 For the wavelet analysis used in this work we chose the Morlet wavelet:

\begin{equation}
\psi_0(t)= \pi^{-\frac{1}{4}}e^{i \omega_0 t} e^{-\frac{t^2}{2}},
\label{vaveleti1}
\end{equation}

\noindent where $\omega_0$ is the non-dimensional frequency and $t$ the non-dimensional time parameter. The wavelet analysis code used is based on the work by Torrence and Compo (1998). The associated Fourier period, P, is $1.03s$ for $\omega_0 = 6$, where $s$ is the wavelet scale (see Table 1 in Torrence and Compo (1998)). One-dimensional wavelet analysis is done for each spatial coordinate. At the beginning and end of the wavelet transforms there are regions where spurious power may arise as a result of the finite extent of the time series. These regions are  referred to as the cone of influence (COI), having a temporal extent equal to the $e$-folding time ($t_d$) of the wavelet function. In our case it is $t_d= \sqrt{2s}=\sqrt{2}\frac{P}{1.03}$. This time scale is the response of the wavelet function to noise spikes and is used in our detection criteria by requiring that reliable oscillations have a duration greater than $t_d$ outside the COI. This imposed a maximum period of $912$s (or $\approx 1$mHz) above which any detected periods were disregarded. The automated wavelet analysis carried out in this work has previously been presented in detail by Bloomfield {\it et al.} (2006). In the current study only the power outside the COI is retained.\par
\begin{figure}
\centering
\includegraphics[width=0.8\textwidth]{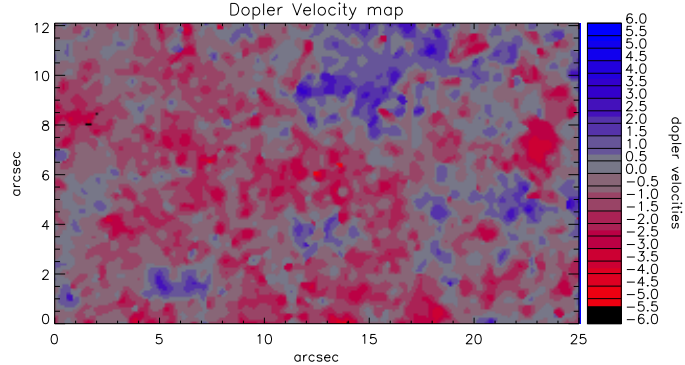}
\caption{ Velocity map from one of the VTT data sets, where the field of view shows only those velocities which appear at the same location as the high frequency oscillations, from the frequency interval $22$mHz to $8$mHz, with power above $15$\% of the maximum level. The remainder is set to a zero value.}
             \label{mapa}
    \end{figure}

  To determine whether registered oscillations are located above down-flows or up-flows, a comparison with velocity maps is performed. In this analysis the power has been normalized and only power in excess of $15$\% of the normalised value is considered (Figure \ref{mapa}.). 

A comparison of the spatial distribution of high frequency oscillations with white-light structures was performed.\par
\begin{figure}
\includegraphics[width=0.8\textwidth]{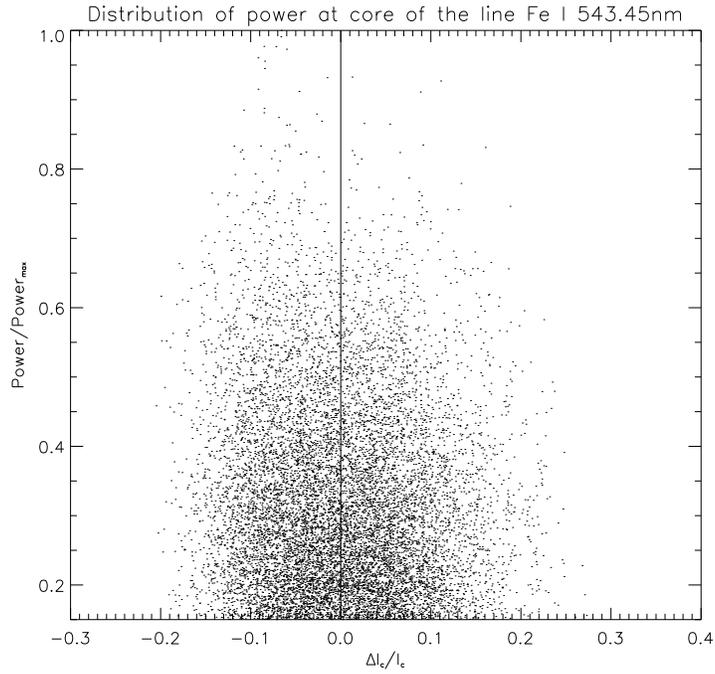}
\caption{~Scatter plot from the DS2data set. High frequency oscillations with power above $15$\% of the maximum level is presented vs. the corresponding pixel intensity in the granulation image. The power was integrated over the frequency range $7$mHz to $17$mHz for this data set. The results from the spectral line $543.45$nm are shown.}
\label{skater1}
\end{figure}

High frequency oscillations power maps are compared pixel by pixel with the corresponding white-light image. In Figure \ref{skater1}. a scatter plot for one frame in the time sequence is shown. The temporal difference between the white-light image and the power map is calculated according to the assumption that the high frequency oscillations travel with the speed of sound.\par
This assumption is oversimplifying the true picture of the lower solar atmosphere and the oscillations within. Up to the present, theoretical models of stellar atmospheres have been based on several simplifying assumptions, such as the restriction to a 1-dimensional (1D) stratification in hydrostatic equilibrium with only a rudimentary parametrization of convection through. Based on such models (e.g. Howleger and M\"uller 1974) it is assumed that in the solar atmosphere the acoustic transit time is approximately 5 minutes (for a height of $2000$km and a sound speed of $7$kms$^{-1})$). For the propagation of acoustic waves in the solar atmosphere over several scale heights, their frequency has to be above the cut-off frequency Equation (\ref{cutoff}).

 \begin{equation}
\omega_{\mathrm{ac}}=\frac{\gamma g}{2 c_s},
 \label{cutoff}
\end{equation}

\noindent where $g$ is the gravitational acceleration, $\gamma=c_p/c_v$ the adiabatic coefficient, and $c_s= \sqrt{H \gamma g}$ with $H$ as density scale height, given as:

\begin{equation}
H \equiv \frac{-\rho_0}{\frac{d \rho_0}{dr}},
\end{equation}

\noindent (Stix (2002), chapter 5.2.4). Therefore in a 1D model the cut-off frequency varies only with atmospheric height. Thus, at a certain height in the atmosphere a reflection layer exists where the values of $g$ and $c_s$ yield the appropriate cut-off frequency. For a given frequency, a wave can be oscillating at one height and be evanescent at another. This situation can cause standing waves for almost the whole range of acoustic frequencies. Fleck (1989) explain that there is a possibility for standing waves, originating from the total reflection of upward propagating waves at the chromosphere - corona transition region. This discovery of standing patterns was confirmed by Espagnet {\it et al.} (1996).\par
Virtually all information about the Sun is deduced from its emergent spectrum. Directly observable manifestations of Solar convection (granulation) thereby influences the emergent spectrum. Also most of our present empirical knowledge of solar surface magnetism stems from the analysis of the Zeeman effect in spectral lines. Routine magnetograms give the impression that cell interiors are non-magnetic, while some recent research showed that's not case (Domingues, 2004; Andjic and Wiehr, 2006; Andic and Vo\'cki\'c, 2007; Trujillo Bueno {\it et al.}, 2004). \par
In light of this information, the assumption that the observed oscillations are purely acoustic is oversimplifying and does not give the full picture of the processes. Never the less, since current theories for the heating of the chromosphere favor acoustic oscillations, part of the current analysis is conducted with this assumption.\par
For this part of the analysis the intergranular lanes (darker areas) were defined as areas where the intensity of the corresponding pixel is less than the modal intensity of the whole field of view.

\begin{figure}
\includegraphics[width=0.8\textwidth]{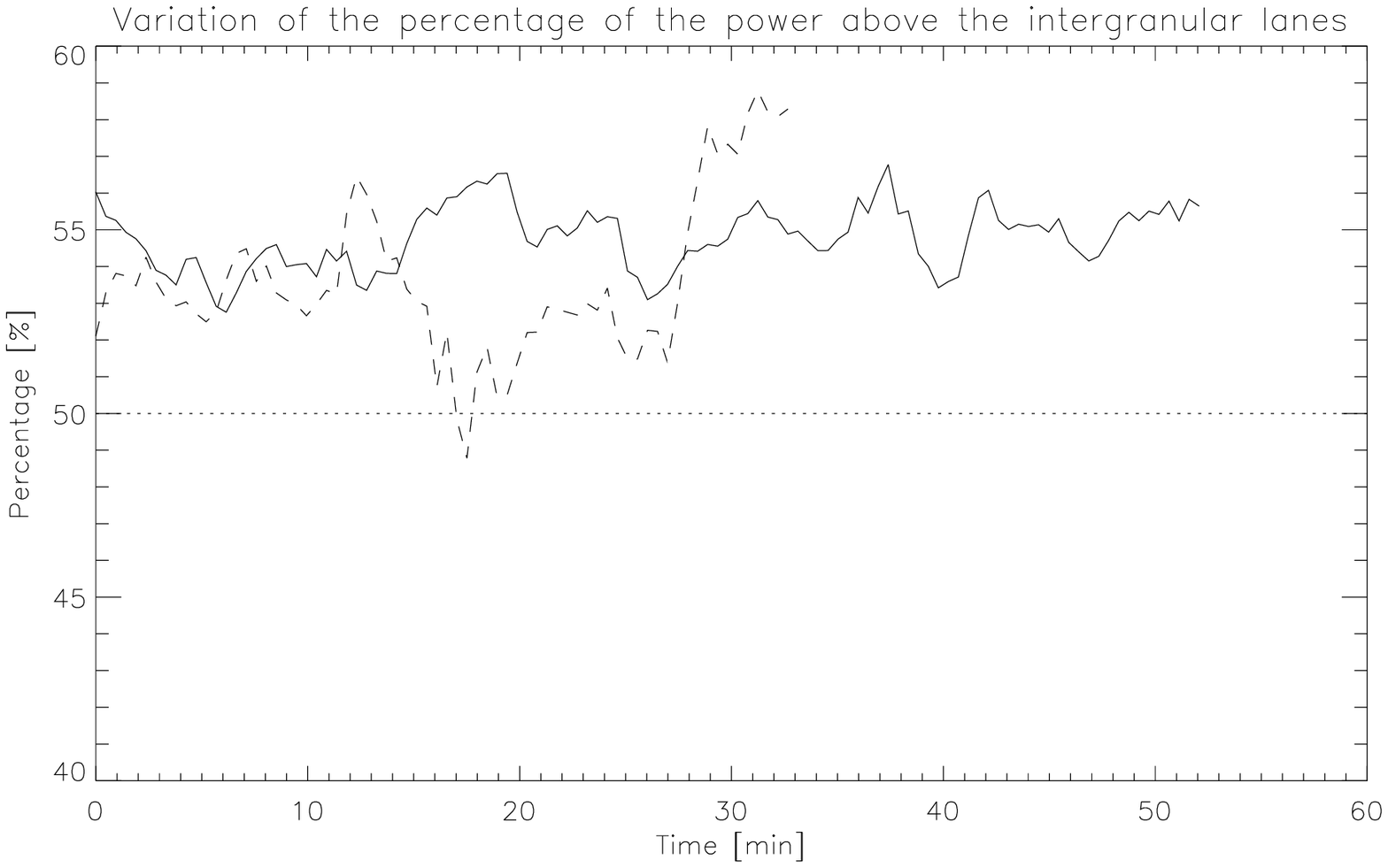}
\caption{Variation of the percentage of observed oscillations located above intergranular lanes from the data sets DS1 and DS2 for the Fe {\sc i} $543.45$nm. The solid line represents the percentage for the frequency range $7$mHz to $17$mHz from the data set DS1. The dashed line represents percentage for data set DS2 and the doted line marks $50$\% level.}
             \label{procentaza}
    \end{figure}

To obtain the percentage of power which is located above dark structures, the power was normalized, and only that power which exceeds $15$\% of the maximum value is considered. This procedure is performed throughut the time sequence yielding the variation of positions with time.  An example is shown in Figure \ref{procentaza}.\par

\section{Results}
\label{results}

Above described analysis is performed for all data sets. Results are presented in Table \ref{tpodaci2}. The $v_B$ present the upward velocities and $v_R$ downward velocities, while $\Sigma v$ presents the sum of all velocities used for this calculation. The formation heights for line cores are calculated for LTE and NLTE ($H_{\mathrm{LTE}}$ and $H_{\mathrm{NLTE}}$) respectively. The percentage is calculated over the whole data set. 

\begin{table}[ht]
\caption{ The percentage of blue and red velocities appearing at the observed locations of high frequency oscillations.}
\label{tpodaci2}       
\begin{tabular}{llllllll}
\hline\noalign{\smallskip}
mark &line[nm]& $v_B$ [\%] & $v_R$ [\%] & $\Sigma v$[km/s]& $H_{LTE}$ &$H_{NLTE}$&power\\
&&&&&[km]&[km]&coverage\\
&&&&&&&[\%]\\
\noalign{\smallskip}\hline\noalign{\smallskip}
DS1 &543.45&32.22& 31.32&-0.0086&660.2&588.7&99.9\\
DS1 &543.29&11.41& 16.27&-0.0508&308.2&258.6&70.7\\
DS2 &543.45& 30.24&29.47&0.0022&660.2&588.7&100\\
DS2 &543.29&8.54&10.75&-0.0204&308.2&258.6&81\\
DS3 &543.45& 7.18& 6.6 & -0.0076&660.2&588.7&72.01\\ 
DS4 & 543.45 & 5.37& 5.94 & 0.0031&660.2&588.7&99\\ 
\noalign{\smallskip}\hline
\end{tabular}
\end{table}

\begin{figure}
\includegraphics[width=0.8\textwidth]{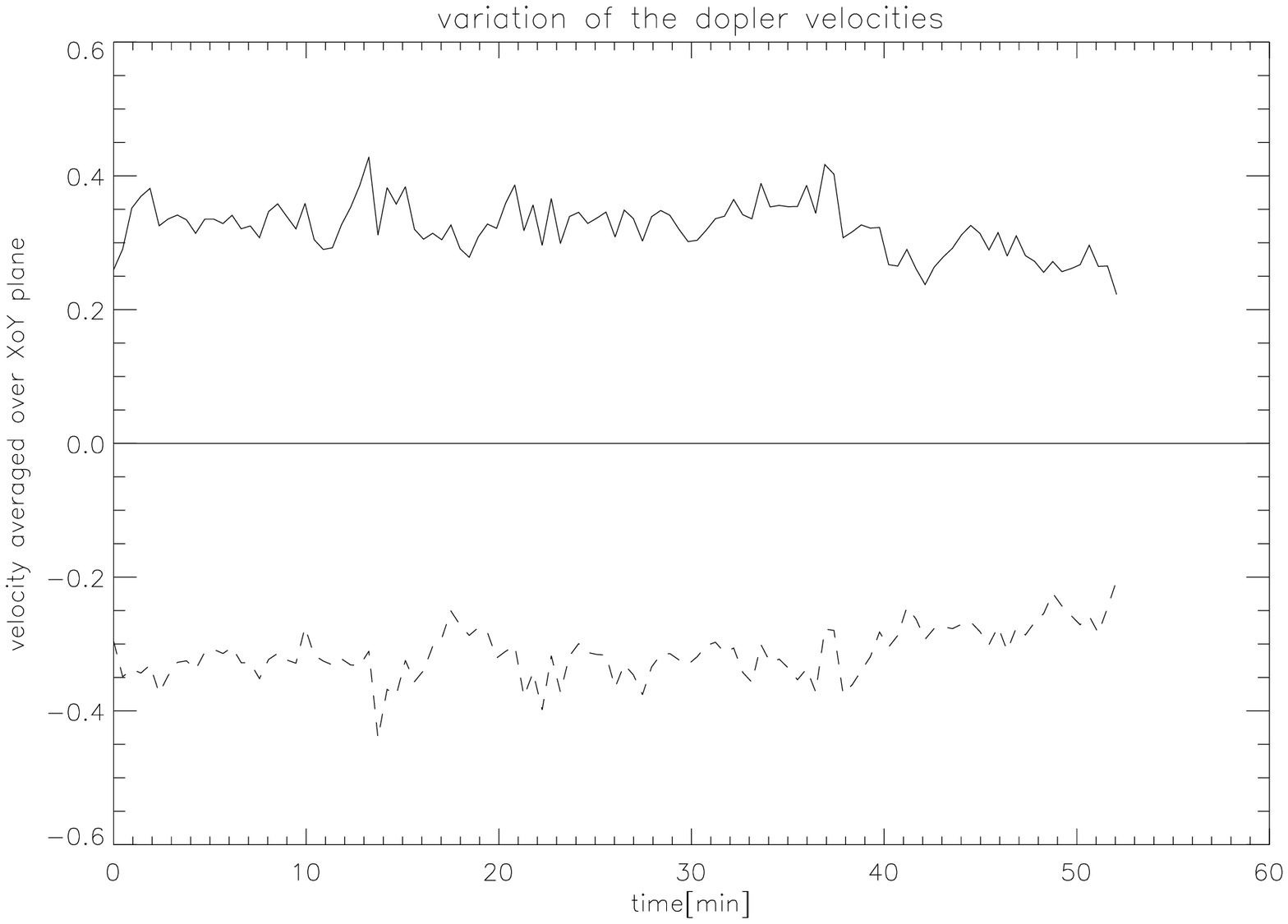}
\caption{Time variation of the percentage of observed velocities located above spatial positions for the observed oscillations. Graph is divided into two parts. The top part shows upward velocities and in bottom part the downward velocities are shown.}
             \label{brzine}
    \end{figure}

The results show in Table \ref{tpodaci2} show that the percentages of upward and downward velocities are similar. This is also visible with results for $\Sigma v$ where values are less than $1$\% of the maximum velocity used in calculations. To see the time variation of this percentage averaging over the field of view is performed. An inspection of Figure \ref{brzine} shows that there is no noticeable trend in the observed velocities. 

\begin{figure}[ht]
\includegraphics[width=0.8\textwidth]{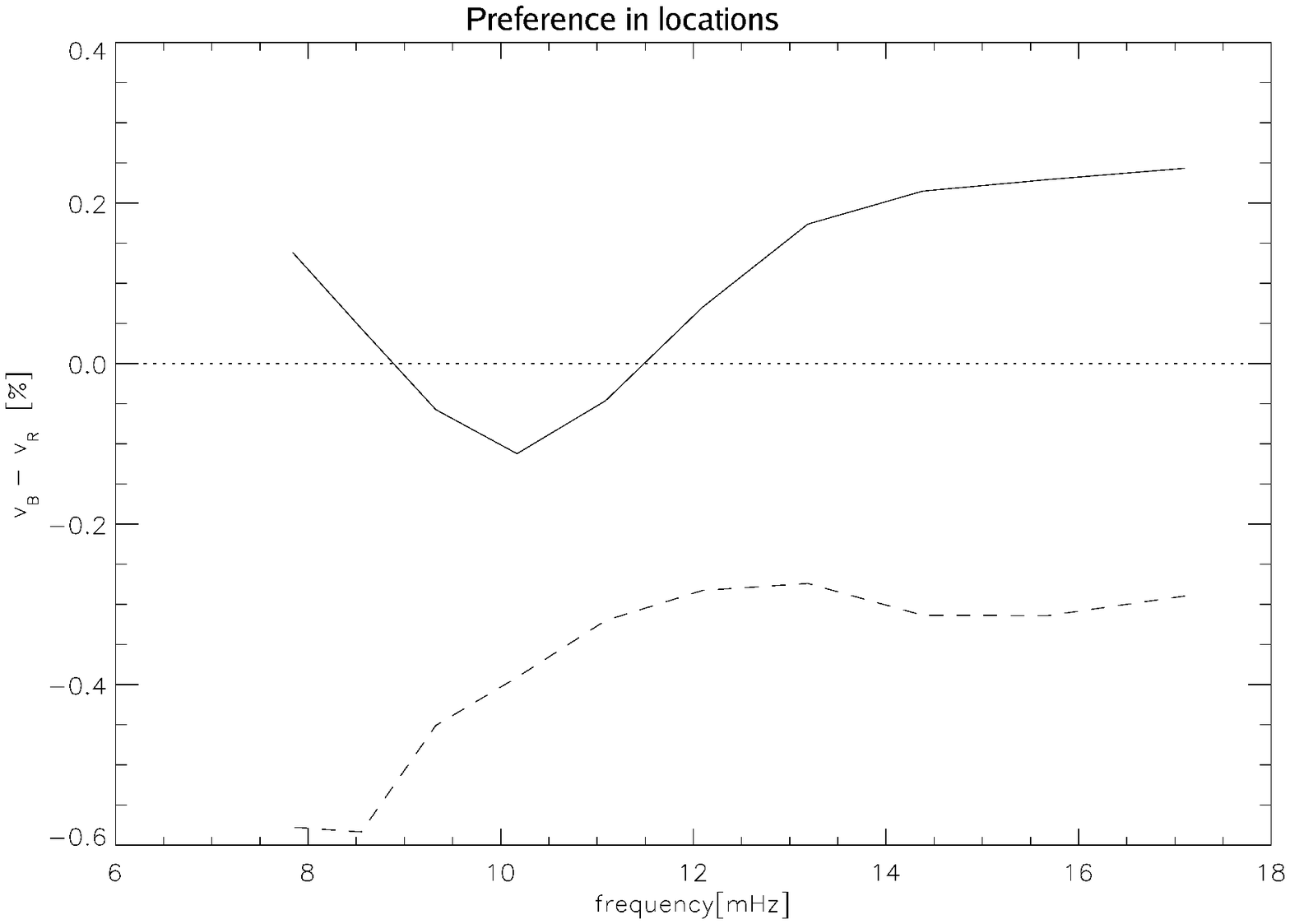}
\caption{Frequency variation of the percentage of observed velocities located above spacial positions of the observed oscillations. The solid line represents the difference between upward and downward velocities ($v_B-v_R$) as a percentage for the core of the $543.45$nm spectral line and dashed line for the $543.29$nm.}
   \label{pozicije}
\end{figure}

Analysis of velocities is done at each frequency. Figure \ref{pozicije} shows that with frequency there is no significant difference in the location of power , since there is only $0.3$\% variation between positions.

Analysis of locations using intensity variations of images taken in the continuum of the lines showed similar results for the spectral line $543.45$nm. 

\begin{figure}
\includegraphics[width=0.8\textwidth]{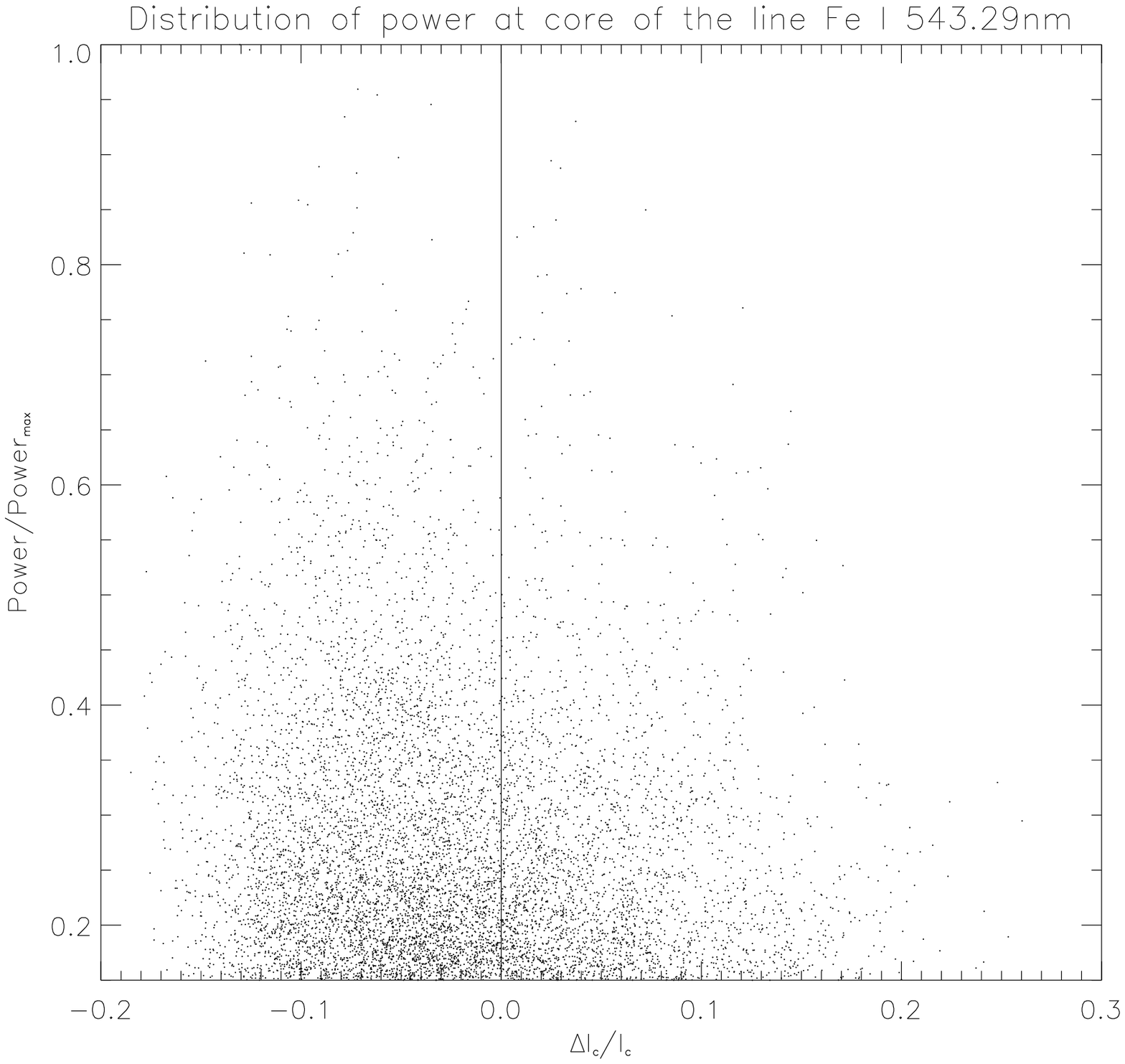}
\caption{~Scatter plot from the DS2 data set for $543.29$nm. High frequency oscillations with power above $15$\% of the maximum level is presented vs. the corresponding pixel intensity in the granulation image. The power was integrated over the frequency range $7$mHz to $17$mHz for this data set.}
   \label{skatervtt}
\end{figure}

\begin{figure}
\includegraphics[width=0.8\textwidth]{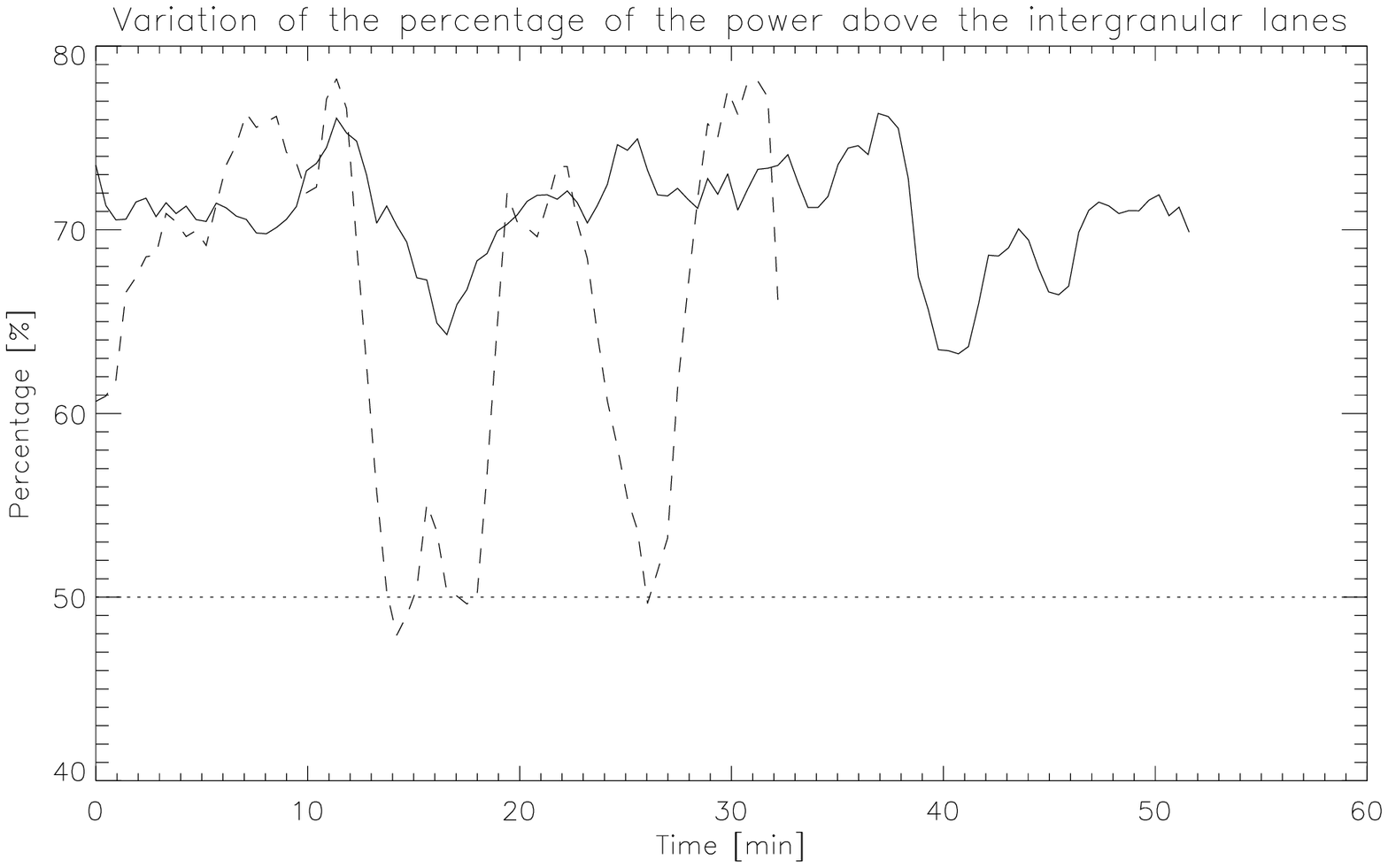}
\caption{Time variation of the percentage of observed oscillations located above intergranular lanes from DS1 and DS2 for $543.29$nm. The solid line represents the percentage for the frequency range $7$mHz to $17$mHz from the data set DS1. The dashed line represents the percentage for the data set DS2 and the dotted line marks $50$\% level.}
             \label{procentaza2}
    \end{figure}

 There is no preference in location of the observed oscillations for the spectral line $543.45$nm, as it can be seen in Figures \ref{skater1} and \ref{procentaza}. Figure \ref{procentaza} shows that the percentage of oscillations above darker areas is around $55$\%, which is not to be taken as certain indicator of a preferable position. Results for the spectral line $543.29$nm show a tendency to appear above darker areas in the continuum image as it can be seen is Figures \ref{skatervtt} and \ref{procentaza2}. In Figure \ref{procentaza2} the percentage of oscillations above darker areas is around $70$\%. Two sudden drops in percentage for the data set DS2 are caused by a decrease in quality of the speckle reconstructed images due to a significant degradation of the seeing condition during observations. 

The percentage of the field of view which contains oscillations is given in Table \ref{tpodaci2}.  This coverage is calculated by summing the positions of all oscillations with power above $15$\% of the maximum power over the complete time sequence. Inspection of the table shows that oscillations observed in the $543.45$nm line, which is formed higher in the atmosphere, cover more of the field of view than the $543.29$nm line. This percentage is lower for DS3. This is because no oscillations were registered over the pore.

\section{Discussion and Conclusions}
\label{conclusions}

The observation used in this work were obtained in quiet solar regions, active regions with G-band structures and an active region containing a pore.  Our findings do not show a dependency on the selected solar locations.  \par
In the data sets from the VTT, the  higher formed  spectral line $543.45$nm show no preference in oscillation site location  (Figures.\ref{skater1} and \ref{procentaza} and Table \ref{tpodaci2}.), while for the  lower formed  line, $543.29$nm, there is a slight preference towards down flow regions (Figures.\ref{skatervtt} and \ref{procentaza2} and Table \ref{tpodaci2}).

\vskip4mm
The influence of atmospheric seeing is mainly visible during speckle reconstruction. The work by von der L\"uhe (1984, 1993, and 1994) estimates an error of $10$\% in intensity introduced by speckle reconstruction. Noise filtering was done following the recommendations of de Boer (1996), so that error levels remain at $10$\%. Seeing influences are most visible with differing data cadences. In the case of good seeing, the cadence was lower than in the case of very good and excellent seeing conditions. Therefore the amount of noise remaining after speckle reconstruction tended to vary only inside predicted error levels (von der L\"uhe, 1984; von der L\"uhe, 1993; von der L\"uhe,1994; de Boer, 1996). A full description of the errors is presented by Andic (2007) \par 

\subsection{Fe {\sc i} $543.45$nm}
Observations (Espagnet {\it et. al} 1996) show that acoustic events occur preferentially in dark intergranular lanes.  Wunnenberg {\it et. al} (2002) claim that high frequency oscillations occur preferentially above down-flows. Observations with the same spectral line as in the work of Wunnenberg {\it et. al} (2002) shows no evidence for this preference. One of the main differences between the work of Wunnenberg {\it et. al} (2002) and the present study has to do with the velocity response functions of the  Fe {\sc i} $543.45$nm line. In this work, velocity response functions were not used.\par
 Figures \ref{skater1} and \ref{procentaza} show that oscillations registered in this line do not show preference in location. \par
Kalkofen (1990) suggests that the location of acoustic oscillations should depend on their frequency. However, our findings do not agree with this prediction. The difference in the location of oscillations of various frequencies is too small to be taken as a serious indicator of their different positions. Figure \ref{pozicije} shows that differences in position with frequency do not exceed $0.3$\%. Also for this line those differences are mainly positive indicating that upward velocities are dominant. Only in the frequency range $9$mHz to $11.5$mHz are there indications for dominance towards donward velocities. Nevertheless, since the difference is less than $1$\% this can not be considered as a significant result.\par
Also, all observed oscillations tend to follow similar patterns, which contradict the statement by Espagnet {\it et. al} (1996) that most energetic oscillations are associated with down-flows in dark areas which are well separated from each other in time and space. Considering the fact that Espagnet {\it et. al} (1996) were working with oscillations of much lower frequencies, this could explain the difference. \par
 The formation height of this spectral line places the core of the line in the vicinity of the temperature minimum. This opens consideration that some of the granulation motions could be registered as oscillations in this work. Espagnet {\it et. al} (1995) show that intensity fluctuations associated with granules disappear in a very short distance, below $100$km above the continuum level. Yet, velocity fluctuations associated with granulation cross the whole thickness of the photosphere. This is typical for larger granules ($2" - 3"$) above which velocities spread throughout the whole photosphere. Therefore it is possible that some of the oscillations detected here from the velocity maps could be originating from granulation. To remove such velocities, without imposing the limit on the registered frequency range, those oscillations with height depended amplitude decay are eliminated. The maximum velocities from averaged granules on this height were noted to be below $1$km/h, while the maximum observed velocities in this work were around $6$km/h, therefore all velocity oscillations with amplitudes below $15$\% were not analysed in this work. This method does not exclude velocities associated to very large granules where the decay is slower.

\subsection{Fe {\sc i} $543.29$nm}
  The slight preference for oscillatory appearance above downwards velocities is found in observations with the this line. Comparing to the intensity of the continuum image shows substantial preference toward darker areas in the granulation pattern.  Figures \ref{skatervtt} and \ref{procentaza2} show that oscillations registered in this line do show preference in their location. \par
 Figure \ref{pozicije} shows that difference in position with frequency does not exceed $0.3$\%. For this line those differences are mainly negative indicating that downward velocities are dominant. But since the difference is less than $1$\%, this also can not be taken as a significant result.\par
 The formation height of this spectral line places the core of the line in the middle of the photosphere. This makes the  results even more sensitive to granulation overshooting. Again, the main concern are velocities which on this level can be, on average, around $1$km/h. Since maximum velocities noted for this height are also around $6$km/h, the method mentioned above cannot successfully remove all velocities which could be associated with granulation. Therefore the error in the results for this line is larger, but no more than $20$\%.

\begin{acknowledgements}
 I acknowledge the Ph.D. scholarship from Max Planck Institute for Solar System Research (Max Planck Institute f\"ur Sonnensystemforschung), Katlenburg-Lindau, Germany.\par 
I wish, especially, to thank Dr. N. Shchukina for calculating the formation heights for the used lines.\par
During the work itself, I had lots of stimulating discussions for which I wish to thank  R. Cameron, A. V\"ogler and E. Wiehr. I wish to thank P. Sutterlin and K. Janssen for giving me the software for the data reduction. For the help with the observations I wish to thank J. Hirzberger and K. Puschmann.\par
I also wish to thank M. Mathioudakis with help in formulating the sentences and together with D.B. Jess for help with the English. 

\end{acknowledgements}

\end{article} 

\begin{thebibliography}{}
\bibitem[\protect\citeauthoryear{Andjic} {2006}]{ja06}Andjic,A. :2006, {\it Serb.Astron.J.} {\bf 172}, 27.
\bibitem[\protect\citeauthoryear{Andjic and Wiehr} {2006}]{andjic06}Andjic,A., Wiehr E. :2006, {\it Publ. Astron. Obs. Belgrade} {\bf 80}, 367.
\bibitem[\protect\citeauthoryear{Andic}{ 2007}]{ja07}Andic,A., {\it Solar Phys.}, accepted.
\bibitem[\protect\citeauthoryear{Andic and Vo\'cki\'c}{ 2007}]{ja07b}Andic,A., Vo\'cki\'c,N. {\it Astron. Astrophys.}, submitted.
\bibitem[\protect\citeauthoryear{Asplund {\it et al.}}{2000}]{asplund00} Asplund,M., Nordlund,\o{A}., Tramperdach,R., Allende Prieto,C., Stein,R.F. :2000, {\it Astron. Astrophys.} {\bf 359},729.
\bibitem[\protect\citeauthoryear{Bendlin, Volkmer, and Kneer}{ 1992}]{bendlin92}Bendlin,C., Volkmer,R., Kneer,F. :1992, {\it Astron. Astrophys.} {\bf 257}, 817.
\bibitem[\protect\citeauthoryear{Bloomfield {\it et al.}}{2006}]{bloomfield06}Bloomfield,D.S., McAteer,R.T.J., Mathioudakis,M., Keenan,F.P. :2006 {\it Astrophys. J.} {\bf 652}, 812.
\bibitem[\protect\citeauthoryear{de Boer, Kneer, and Nesis}{1992}]{boer92}de Boer,C.R., Kneer,F., Nesis,A. :1992, {\it Astron. Astrophys.} {\bf 257}, L4.
\bibitem[\protect\citeauthoryear{de Boer and Kneer}{1994}]{boer94}de Boer,C.R., Kneer,F. :1994, In Robertson,J.G., Tango,W.J. (eds.),{\it Very High Angular Resolution Imaging}, {\it IAU Symp.} {\bf 158}, 398.
\bibitem[\protect\citeauthoryear{de Boer}{1996}]{boer96}de Boer,C.R. :1996, {\it Astron. Astrophys. Suppl.} {\bf 120}, 195.
\bibitem[\protect\citeauthoryear{Deubner and Laufer}{1983}]{deubner83}Deubner,F.L., Laufer,J. :1983, {\it Solar Phys.} {\bf 82}, 151.
\bibitem[\protect\citeauthoryear{Dom\`\i nguez}{2004}]{ita}Dom\`\i nguez,I.F. :2004, {\it Quiet Sun Magnetic Fields}, Copernikus GmbH, Katlenburg-Lindau, Germany, p.111.
\bibitem[\protect\citeauthoryear{Espagnet {\it et al.}}{1996}]{espagnet96}, Espagnet, O. {\it et al.}: 1996, {\it Astron. Astrophys.} {\bf 313}, 297.
\bibitem[\protect\citeauthoryear{Espagnet {\it et al.}}{1995}]{espagnet95}Espagnet, O., Muller, R., Roudier, Th., Mein, N., Mein, P.: 1995, {\it Astron. Astrophys. Suppl.}{\bf 109}, 79.
\bibitem[\protect\citeauthoryear{Fleck and Deubner}{1989}]{fleck89}Fleck, B., Deubner, F.L.: 1989, {\it Astron. Astrophys.} {\bf224}, 245.
\bibitem[\protect\citeauthoryear{Holweger and M\"uller}{1974}]{holweger74}Holweger,H., M\"uller,E.A.: 1974, {\it Solar Phys.} {\bf 39}, 19.
\bibitem[\protect\citeauthoryear{Janssen}{2003}]{katja}Janssen,K.: 2003, {\it Struktur und Dynamik kleinskaliger Magnetfelder der Sonnenatmosphaere}, Copernikus GmbH, 2003, Kaltenburg-Lindau, Germany, p.123.
\bibitem[\protect\citeauthoryear{Kalkofen}{1990}]{kalkofen90}Kalkofen,W.: 1990,In: Priest, E.R., Krishan, V. (eds.),{\it Basic Plasma Processes on the Sun},{\it IAU Symp.} {\bf 142}, 197.
\bibitem[\protect\citeauthoryear{Kalkofen}{2001}]{kalkofen01}Kalkofen,W.: 2001, {\it Astrophys. J.} {\bf 557}, 376.
\bibitem[\protect\citeauthoryear{Keller and von der L\"uhe}{1992}]{keller92}Keller,C.U., von der L\"uhe,O.: 1992,  {\it Astron. Astrophys.} {\bf 261}, 321.
\bibitem[\protect\citeauthoryear{Koschinsky, Kneer, and Hirzberger}{2001}]{koschinsky01}Koschinsky,M., Kneer,F., Hirzberger,J. :2001, {\it Astron. Astrophys.} {\bf 365}, 588.
\bibitem[\protect\citeauthoryear{Lighthill}{1951}]{lighthill51}Lighthill,M.J.: 1951, {\it Proceedings of the Royal Society}  {\bf A 211}, 564.
\bibitem[\protect\citeauthoryear{Proudman}{1952}]{proudman52}Proudman,I. :1952{\it Proc. Roy. Soc. London} {\bf A 214}, 119.
\bibitem[\protect\citeauthoryear{Shchukina and Trujillo Bueno}{2001}]{shukina01}Shchukina,N.G., Trujillo Bueno,J.: 2001, {\it Astrophys. J.} {\bf 550}, 970.
\bibitem[\protect\citeauthoryear{Stein}{1967}]{stein67}Stein,R.F.: 1967, {\it Solar Phys.} {\bf 2}, 385.
\bibitem[\protect\citeauthoryear{Stix}{2002}]{stix02}Stix, M.,: 2002, {\it The Sun an introduction}, Springer -Verlag Berlin Heidelberg.
\bibitem[\protect\citeauthoryear{Torrence and Compo}{1998}]{torrence97}Torrence,C., Compo,G.P.: 1998, {\it Bull. Amer. Meteor. Soc.} {\bf 79}, 61.
\bibitem[\protect\citeauthoryear{Trujillo Bueno, Shchukina, and Asensio Ramos}{2004}]{trujillo04}Trujillo Bueno,J. ,Shchukina,N. ,Asensio Ramos,A.:2004, {\it Nature} {\bf 430}, 326.
\bibitem[\protect\citeauthoryear{von der L\"uhe}{1984}]{luhe84}von der L\"uhe,O. :1984, {\it J.Opt.Soc.Am A} {\bf 1}, 510.
\bibitem[\protect\citeauthoryear{von der L\"uhe}{1994a}]{luhe94}von der L\"uhe,O. :1994a, {\it Astron. Astrophys.} {\bf 281}, 889.
\bibitem[\protect\citeauthoryear{von der L\"uhe}{1994b}]{luhe93}von der L\"uhe,O. :1994b, {\it Astron. Astrophys.} {\bf 268}, 347.
\bibitem[\protect\citeauthoryear{Wunnenberg, Kneer, and Hirzberger}{2002}]{maren}Wunnenberg,M., Kneer,F., Hirzberger,J.: 2002, {\it Astron. Astrophys.} {\bf 395}, L51.
\end{thebibliography}
\end{document}